\documentclass[10pt,conference]{IEEEtran}
\usepackage{cite}
\usepackage{graphicx}
\usepackage{epsfig,psfrag}
\usepackage{balance}
\usepackage{array}
\usepackage{theorem}
\usepackage{amsmath}
\usepackage{amssymb}
\usepackage{flushend}
\usepackage{tikz}

\newtheorem{theorem}{Theorem}
\newtheorem{fact}{Fact}

\newcommand{\mc}{\mathcal}
\newcommand{\mb}{\mathbf}

\newcommand{\figurewidth}{\linewidth}

\begin{document}

\title{Results on the Redundancy of Universal Compression for Finite-Length Sequences}

\author{Ahmad Beirami and Faramarz Fekri\\School of Electrical and Computer Engineering\\Georgia Institute of Technology,~Atlanta~GA~30332, USA\\Email:~\{beirami,~fekri\}@ece.gatech.edu\vspace{-0.2in}}

\maketitle

\begin{abstract}
In this paper, we investigate the redundancy of universal coding schemes on  smooth parametric sources in the {\em finite}-length regime.
We derive an upper bound on the probability of the event that a sequence of length $n$, chosen using Jeffreys' prior from the family of parametric sources with $d$ unknown parameters, is compressed with a redundancy smaller than $(1-\epsilon)\frac{d}{2}\log n$ for any $\epsilon>0$.
Our results also confirm that for large enough $n$ and $d$, the average minimax redundancy provides a good estimate for the redundancy of most sources.
Our result may be used to evaluate the performance of universal source coding schemes on finite-length sequences.
Additionally, we precisely characterize the minimax redundancy for two--stage codes. We demonstrate that the two--stage assumption incurs a negligible redundancy especially when the number of source parameters is large. Finally, we show that the redundancy is significant in the compression of small sequences.
\end{abstract}


\IEEEpeerreviewmaketitle

\section{Introduction}
\label{sec:intro}
Recently, there has been a tremendous increase in the amount of data being stored in the storage systems. The redundancy present in the data may be leveraged to significantly reduce the cost of data maintenance as well as data transmission.
 In many cases, however, the data consists of several small files that need to be compressed and retrieved individually, i.e., a {\em finite}-length compression problem. Moreover, different data sets may be of various natures, hence little a priori assumptions may be made regarding the probability distribution of the data, i.e., universal compression.
This necessitates the study of the universal compression of finite-length sequences.

 In this paper, we investigate the universal compression of smooth parametric sources.
Denote $\mc{A}$ as a finite alphabet. Let $C_n:\mc{A}^n \to \{0,1\}^*$ be an injective mapping from the set $\mc{A}^n$ of the sequences of length $n$ over $\mc{A}$ to the set $\{0,1\}^*$ of binary sequences.
We use the notation $x^n = (x_1,...,x_n) \in \mc{A}^n$ to present a sequence of length $n$.
Let $\theta = (\theta_1,...,\theta_d)$ be a $d$-dimensional parameter vector.
Denote $\mu_\theta$ as the parametric information source with $d$ unknown parameters, where $\mu_\theta$ defines a probability measure on any sequence $x^n \in \mc{A}^n$.
Denote $\mc{P}^d$ as the \emph{family} of sources with $d$-dimensional unknown parameter vector $\theta$.
 Let $H_n(\theta)$ be the source entropy given parameter vector $\theta$, i.e.,
\begin{equation}
H_n(\theta) =  \mb{E} \log \left(\frac{1}{\mu_\theta(X^n)} \right) =\sum_{x^n}\mu_{\theta}(x^n) \log\left(\frac{1}{\mu_\theta(x^n)}\right).\footnote{Throughout this paper all expectations are taken with respect to the true unknown parameter vector~$\theta$.}
\end{equation}
In this paper $\log(x)$ always denotes the logarithm of $x$ in base $2$.
Let $l(C_n, x^n) = l_n(x^n)$  denote the length function that describes the codeword length associated with the sequence $x^n$. Denote $L_n$ as the set of all regular length functions on an input sequence of length $n$.

Denote $R_n(l_n,\theta)$ as the expected redundancy of the code on a sequence of length $n$, defined as the difference between the expected codeword length and the entropy,
\begin{equation}
R_n(l_n,\theta)  = \mb{E}l_n(X^n) -  H_n(\theta).
\label{eq:def_redundancy}
\end{equation}
The expected redundancy is always non-negative. For a code that asymptotically achieves the entropy rate with length function $l_n$, $\frac{1}{n} R_n(l_n,\theta)\to 0$ as $n \to \infty$ for all $\theta$.
The maximum expected redundancy for a length function of a code with length function $l_n$ is given as $
R_n(l_n) = \max_{\theta \in \Theta^d} R_n(l_n,\theta),$
which may be minimized over all codes to achieve the minimax expected redundancy~\cite{Clarke_Barron, Minimax_Redundancy_Memoryless, Davisson_source_matching}
\begin{equation}
\bar{R}_n = \min_{l_n \in L_n} \max_{\theta \in \Theta^d} R_n(l_n,\theta).
\end{equation}
 The leading term in the average minimax redundancy is asymptotically $\frac{d }{2}\log n$. Rissanen demonstrated that for the universal compression of the family $\mc{P}^d$ of the parametric sources with parameter vetcor $\theta$, the redundancy of the codes with \emph{regular} length functions $l_n$ is asymptotically lower bounded by $R_n(l_n,\theta) \geq (1 -  \epsilon)\frac{d }{2}\log n$~\cite{Rissanen_1984,Rissanen_Complexity, Rissanen_stochastic_complexity}, for all $\epsilon>0$ and almost all sources.  This asymptotic lower bound is tight since there exist coding schemes that achieve the bound asymptotically~\cite{Rissanen_Complexity, CTW95}. This result was later extended in~\cite{Merhav_Feder_DCC,Feder_Hierarchical,redundancy_individual} to more general classes of sources.
In~\cite{Allerton10}, we extended Rissanen's probabilistic treatment of redundancy to  the universal compression in {\em finite}-length memoryless sources for the family of two--stage codes. However, the two--stage code assumption is restrictive and incurs an extra redundancy.

 In this paper,  we extend our previous work to the family of parametric sources. We also relax the two--stage codeing constraint by considering conditional two--stage codes so that the coding scheme is optimal in the sense that it achieves the minimax redundancy. Further, we derive the extra redundancy incurred of two--stage codes. The rest of this paper is organized as follows. In Section~\ref{sec:background}, after a review of the previous work, we formally state the problem of redundancy for finite-length universal compression of parametric sources.
In Section~\ref{sec:two_part_minimax}, we present our main results on the compression of conditional two--stage and two--stage codes.   In Section~\ref{sec:results}, we demonstrate the significance of our results through several examples.

\section{Background Review and Problem Statement}
\label{sec:background}
In this section, after a brief review of the previous work, we state the finite-length redundancy problem.
Let $l_n^{\theta}$ denote the (non-universal) length function induced by a parameter $\theta \in \Theta^d$.
Denote $l_n$ as the length function on the input sequence of length $n$. Denote $R_n(l_n,\theta)$ as the expected redundancy of the universal compression of source $\mu_\theta\in \mc{P}^d$ using the length function $l_n$.
Let $I_n(\theta)$ be the Fisher information matrix for parameter vector $\theta$ and a sequence of length $n$,
\begin{equation}
I_n(\theta) = \{I_n^{ij}(\theta)\} = \frac{1}{n\log e}\mb{E}\left\{\frac{\partial^2}{\partial \theta_i \partial \theta_j} \log\left( \frac{1}{\mu_\theta(X^n)}\right) \right\}.
\end{equation}
Fisher information matrix quantifies the amount of information, on the average, that each symbol in a sample sequence of length $n$ from the source conveys about the source parameters.

In this paper, we assume that the following conditions hold:
\begin{enumerate}
\item $\Theta^d$ forms a compact set.
\item  $\lim_{n\to\infty} I_n(\theta)$ exists and the limit is denoted by $I(\theta)$.
\item All elements of the Fisher information matrix $I_n(\theta)$ are continuous in $\Theta^d$.
\item $\int_{\Theta^d} |I(\theta)|^{\frac{1}{2}}d\theta < \infty.$
\item The family $\mc{P}^d$ has a minimal representation with the $d$-dimensional parameter vector~$\theta$.
\end{enumerate}

Rissanen proved an asymptotic lower bound on the universal compression of an information sources with $d$ parameters as~\cite{Rissanen_Complexity, Rissanen_stochastic_complexity}:
\begin{fact}
For all parameters $\theta$, except in a set of asymptotically Lebesgue volume zero, we have
\begin{equation}
\lim_{n\to\infty} \frac{R_n(l_n,\theta)}{\frac{d}{2}\log n} \geq 1-\epsilon,~~\forall \epsilon>0.
\end{equation}
\label{thm:Rissanen}
\vspace{-10pt}
\end{fact}
While Fact~\ref{thm:Rissanen} describes the asymptotic fundamental limits of the universal compression of parametric sources, it does not provide much insight for the case of {\em finite}-length $n$. Moreover, the result excludes an asymptotically volume zero set of parameter vectors $\theta$ that has non-zero volume for any finite $n$.

In~\cite{Clarke_Barron}, Clarke and Barron derived the expected minimax redundancy $\bar{R}_n$ for memoryless sources, later generalized in~\cite{atteson_markov} by Atteson for Markov sources, as the following:
\begin{fact}
The average minimax redundancy is asymptotically given by
\begin{equation}
\bar{R}_n = \frac{d}{2}  \log\left( \frac{n}{2\pi} \right) + \log \int
|I_n(\theta)|^{\frac{1}{2}}d\theta + O\left(\frac{1}{n}\right).
\label{eq:minimax}
\end{equation}
\vspace{-10pt}
\end{fact}
The average minimax redundancy characterizes the maximum redundancy over the space $\Theta^d$ of the parameter vectors. However, it does not say much about the rest of the space of the parameter vectors.
It is known that if $\mu_\theta(x^n)$ is a measurable function of $\theta$ for all $x^n$, the average minimax redundancy is equal to the capacity of the channel between the parameter vector $\theta$ and the sequence $ x^n$, i.e., $\bar{R}_n = \sup_p I_p(\Theta; X^n)$, where $p(\cdot)$ is a probability measure on the space of the parameter vector $\theta$~\cite{Davisson_noiseless_coding,Merhav_Feder_DCC}.
The average minimax redundancy is obtained when the parameter vector $\theta$ follows the capacity achieving prior, which is Jeffreys' prior in the case of parametric sources. Jeffreys' prior is given by~\cite{Minimax_Redundancy_Memoryless}
\begin{equation}
p(\theta) = \frac{|I(\theta)|^{\frac{1}{2}}}{\int|I(\lambda)|^\frac{1}{2}d\lambda}.
\end{equation}

In a two--stage code, to encode the sequence $x^n$ the compression scheme attributes $m$ bits  to identify an estimate for the unknown source parameters. Then, in the second stage of the compression, it is assumed that the source with the estimated parameter has generated the sequence. In this case, there will be $2^m$ possible estimate points in the parameter space for the identification of the source. Let $\Phi^m = \left\{ \phi_1,...,\phi_{2^m}\right\}$ denote the set of all estimate points with an $m$-bit estimation budget. Note that for all $i$, we have $\phi_i \in \Theta^d$~\cite{Rissanen_two-part, Grunwald_book}.

Denote $l_n^{2p}$ as the two--stage length function for the compression of sequences of length $n$. For each sequence $x^n$, there exists an estimate point in the set of the estimate points, i.e., $\gamma = \gamma(x^n,m) \in \Phi^m$, which is optimal in the sense that it minimizes the code length and the average redundancy. In other words, $\gamma$ is the maximum likelihood estimation of the unknown parameter in the set of the estimate parameters, i.e.,
\begin{equation}
\gamma = \arg\min_{\phi_i \in \Phi^m} \log\left( \frac{1}{\mu_{\phi_i}(x^n)} \right)= \arg \max_{\phi_i \in \Phi^m} \mu_{\phi_i}(x^n).
\end{equation}
The two--stage universal length function for the sequence $x^n$ is then given by
\vspace{-10pt}
\begin{equation}
l_n^{2p}(x^n) = m + l_n^{\gamma}(x^n),
\label{eq:length_func}
\end{equation}
where $l_n^{\gamma}$ denotes the length function induced by the parameter $\gamma \in \Phi^m$. Let $L_n^{2p}$ be the set of all two--stage codes that could be described as in~(\ref{eq:length_func}). Further denote $\mu_{\gamma}(x^n)$ as the probability measure induced by $\gamma$.

Increasing the bit budget $m$ for the identification of the unknown source parameters results in an exponential growth in the number of estimate points, and hence, smaller $l_n^\gamma(x^n)$ on the average  due to the more accurate estimation of the unknown source parameter vector. On the other hand, $m$ directly appears as part of the compression overhead in~(\ref{eq:length_func}). Therefore, it is desirable to find the optimal $m$ that minimizes the total expected codeword length, which is $\mb{E}l_n^{2p}(X^n) =  m+ \mb{E}l_n^{\gamma}(X^n)$.

In this paper, we ignore the redundancy due to the integer constraint on the length function. Thus, we use the Shannon code for each estimated parameter to bound the average redundancy of two--stage codes. Thus, ignoring the integer constraint on the codeword length we have
\begin{equation}
R_n(l_n^{2p},\theta) = m+ \mb{E}\log \left( \frac{1}{\mu_{\gamma}(X^n)}\right) - H_n(\theta).
\label{eq:problem_statement}
\end{equation}
Further, let $\bar{R}_n^{2p}$ denote the average minimax redundancy of the two--stage codes, i.e.,
\begin{equation}
\bar{R}_n^{2p} = \min_{l_n^{2p} \in L_n^{2p}} \max_{\theta \in \Theta^d} R_n(l_n^{2p}, \theta).
\end{equation}

In a two--stage code, we already have some knowledge about the sequence $x^n$ through the optimally estimated parameter $\gamma(x^n)$ (maximal likelihood estimation) that can be leveraged for encoding $x^n$ using the length function $l_n^\gamma(x^n)$.
The two--stage length function in~(\ref{eq:length_func}) defines an incomplete coding length, which does not achieve the equality in Kraft's inequality. Thus, it is not optimal in the sense that it does not achieve the optimal compression among all length functions. Further, it does not achieve the average minimax redundancy~\cite{Allerton10,Grunwald_book}. Conditioned on $\gamma(x^n)$, the length of the codeword for $x^n$ may be further decreased~\cite{Rissanen_two-part}.

Let $S_m(\gamma)$ be the collection of all $x^n$ for which the optimally estimated parameter is $\gamma$, i.e.,
\begin{equation}
S_m(\gamma) \triangleq \left\{ x^n \in \mc{A}^n:~ \mu_{\gamma}(x^n) \geq \mu_{\phi_i}(x^n) ~~\forall \phi_i \in \Phi^m \right\}.
\end{equation}
Further, let $A_m(\gamma)$ denote the total probability measure of all sequences in the set $S_m(\gamma)$, i.e.,
\begin{equation}
A_m(\gamma) = \sum_{x^n \in S_m(\gamma)} \mu_\gamma(x^n).
\label{eq:A_m}
\end{equation}
Thus, the knowledge of $\gamma(x^n)$ in fact changes the probability distribution of the sequence.
Denote $\mu_\gamma(x^n|x^n \in S_m(\gamma))$ as the conditional probability measure of $x^n$ given $\gamma$ is known to be such that $x^n \in S_m(\gamma)$, i.e., the probability distribution that is normalized to $A_m(\gamma)$. That is
\begin{equation}
\mu_\gamma(x^n|x^n \in S_m(\gamma)) = \frac{\mu_\gamma(x^n)}{A_m(\gamma)}.
\end{equation}
Note that $\mu_\gamma(x^n|x^n \in S_m(\gamma)) \geq \mu_\gamma(x^n)$ due to the fact that $A_m(\gamma) \leq 1$.
Let $l_n^\gamma(x^n|x^n \in S_m(\gamma))$ be the codeword length corresponding to the conditional probability distribution, which is decreased to $\mb{E} \log \left(\frac{A_m(\gamma(X^n))}{\mu_\gamma(X^n)}\right)$.
Denote $l_n^{c2p}$ as the conditional two--stage length function for the compression of sequences of length $n$ using the normalized maximum likelihood, which is given by
\begin{equation}
l_n^{c2p} = m + l_n^\gamma(x^n|x^n \in S_m(\gamma)).
\label{eq:length_conditional}
\end{equation}
Therefore, the average redundancy of the conditional two--stage scheme is given by
\begin{equation}
R_n(l_n^{c2p},\theta) =  m + \mb{E} \log \left( \frac{A_m(\gamma (X^n))}{\mu_\gamma(X^n)}\right) - H_n(\theta).
\label{eq:problem_statement_cond}
\end{equation}

Denote $L_n^{c2p}$ as the set of the conditional two--stage codes that are described using~(\ref{eq:length_conditional}).
Let $\bar{R}_n^{c2p}$ denote the average minimax redundancy of the conditional two--stage codes, i.e.,
\begin{equation}
\bar{R}_n^{c2p} = \min_{l_n^{c2p} \in L_n^{c2p}} \max_{\theta \in \Theta^d} R_n(l_n^{c2p}, \theta).
\end{equation}
 Rissanen demonstrated that this conditional version of two--stage codes is in fact optimal in the sense that it achieves the average minimax redundancy~\cite{Rissanen_fisher_information}.
In other words, $\bar{R}_n^{c2p} = \bar{R}_n$,
where $\bar{R}_n$ is the average minimax redundancy in~(\ref{eq:minimax}).

\section{Main Results on the Redundancy}
\label{sec:two_part_minimax}
In this section, we present our main results on the compression of parametric sources. The proofs are omitted due to the lack of space.
We derive a lower bound on the probability of the event that a parametric source $P$ is compressed with redundancy greater than the redundancy level $R_0$, i.e., $\mb{P}[R_n(l_n,\theta)>R_0]$. This bound demonstrates the fundamental limits of the universal compression for finite-length $n$. The following is our main result:
\begin{theorem}
Assume that the parameter vector $\theta$ follows Jeffreys' prior in the universal compression of the family of parametric sources $\mc{P}^d$.
Let $\epsilon$ be a real number. Then,
\begin{equation}
\mb{P}\left[ \frac{R_n(l_n^{c2p},\theta)}{\frac{d}{2}\log n} \geq 1- \epsilon \right] \geq 1 - \frac{1}{\int |I(\theta)|^{\frac{1}{2}}d \theta} \left(\frac{2 \pi}{n^\epsilon} \right)^{\frac{d}{2}}.
\end{equation}
\label{thm:Main_general}
\vspace{-5pt}
\end{theorem}
This theorem is derived for the conditional two--stage length functions. Note that Fact~\ref{thm:Rissanen} is readily deduced from Theorem~\ref{thm:Main_general} by letting $n \to \infty$.

Next, we characterize the redundancy of two--stage codes. Let $l_n^2p$ be the two--stage length function as defined in~(\ref{eq:length_func}). Further, denote $R_n^2p(l_n^{2p},\theta)$ as
the expected redundancy of the universal compression for the source $P\in \mc{P}^d$ with parameter vector $\theta$ using $l_n^{2p}$.
The following theorem sets a lower bound on the redundancy of two--stage codes.
\begin{theorem}
Consider the universal compression of the family of parametric sources $\mc{P}^d$ with the parameter vector $\theta$ that follows Jeffreys' prior.
Let $\epsilon$ be a real number. Then,
\begin{equation}
\mb{P}\left[ \frac{R_n(l_n^{2p},\theta)}{\frac{d}{2}\log n} \geq 1-\epsilon \right] \geq 1 - \frac{C_d}{\int|I(\theta)|^\frac{1}{2}d\theta} \left( \frac{d}{en^{\epsilon}}\right)^{\frac{d}{2}},
\end{equation}
where $C_d$ is the volume of the $d$-dimensional unit ball, which is
\vspace{-5pt}
\begin{equation}
C_d = \frac{\Gamma\left(\frac{1}{2}\right)^d}{\Gamma\left( \frac{d}{2} + 1\right)}.
\label{eq:unit_ball}
\end{equation}
\label{thm:Main}
\vspace{-5pt}
\end{theorem}

Further, we precisely characterize the extra redundancy due to the two--stage assumption on the code as follows:
\begin{theorem}
In the universal compression of the family of parametric sources $\mc{P}^d$, the average minimax redundancy of two--stage codes is obtained by
\begin{equation}
\bar{R}^{2p}_{n} = \bar{R}_n + g(d) +O\left(\frac{1}{n}\right).
\label{eq:minimax_two_part}
\end{equation}
Here, $\bar{R}_n$ is the average minimax redundancy defined in~(\ref{eq:minimax}) and $g(d)$ is the two--stage penalty term given by
\begin{equation}
g(d) = \log\Gamma\left(\frac{d}{2}+1\right) - \frac{d}{2}
\log\left(\frac{d}{2e}\right).
\label{eq:extra_redundancy}
\end{equation}
\label{thm:minimax}
\vspace{-10pt}
\end{theorem}

\section{Elaboration on the Results}
\label{sec:results}
In this section, we elaborate on the significance of our results. In Section~\ref{subsec:performance_small}, we demonstrate that the average minimax redundancy underestimates the performance of source coding in the small to moderate length $n$ for sources with small $d$.  In Section~\ref{subsec:comparison}, we compare the performance of two--stage codes with conditional two--stage codes. We show that the penalty term of two--stage coding is negligible for sources with large $d$ as well as for the sequences of long $n$.
In Section~\ref{subsec:performance_large}, we demonstrate that as the number of source parameters grow, the minimax redundancy well estimates the performance of the source coding.

\begin{figure}[tb]
\centering
\vspace{-0.05in}
\psfrag{ylabel}{$R_0$}
\psfrag{xlabel}{$P_0$}
\psfrag{aaaaaaaaaaaaaaaaaaaa}{\scriptsize{\hspace{-2pt}$\mb{P}[R_n(l_n^{c2p},\theta) \geq R_0] \geq P_0$}}
\psfrag{dataaaaaaaaaaaaaaaaa1}{\footnotesize{$n = 8$ (c2p)}}
\psfrag{data2}{\footnotesize{$n = 8$ (Minimax)}}
\psfrag{data3}{\footnotesize{$n = 32$ (c2p)}}
\psfrag{data4}{\footnotesize{$n = 32$ (Minimax)}}
\psfrag{data5}{\footnotesize{$n = 128$ (c2p)}}
\psfrag{data6}{\footnotesize{$n = 128$ (Minimax)}}
\psfrag{data7}{\footnotesize{$n = 512$ (c2p)}}
\psfrag{data8}{\footnotesize{$n = 512$ (Minimax)}}
\epsfig{width=\figurewidth,file=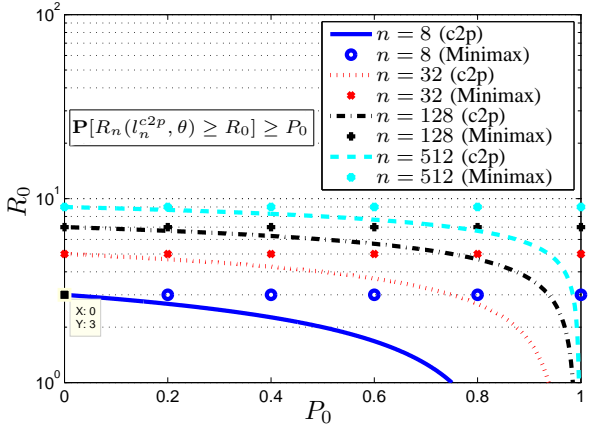}
\vspace{-0.15in}
\vspace{-5pt}
\caption{\small{Average redundancy of the conditional two--stage codes ({\em c2p}) and the average minimax redundancy ({\em Minimax}) as a function of the fraction of sources $P_0$ with $R_n(l_n^{c2p},\theta)>R_0$. Memoryless source $\mc{M}_0^3$ with $k=3$ and $d=2$.}}
\label{fig:M0_3}
\end{figure}

\begin{figure}[tb]
\centering
\vspace{-0.05in}
\psfrag{ylabel}{$R_0$}
\psfrag{xlabel}{$P_0$}
\psfrag{aaaaaaaaaaaaaaaaaaaa}{\scriptsize{\hspace{-2pt}$\mb{P}[R_n(l_n^{c2p},\theta) \geq R_0] \geq P_0$}}
\psfrag{dataaaaaaaaaaaaaaaaa1}{\footnotesize{$n = 12$ (c2p)}}
\psfrag{data2}{\footnotesize{$n = 12$ (Minimax)}}
\psfrag{data3}{\footnotesize{$n = 50$ (c2p)}}
\psfrag{data4}{\footnotesize{$n = 50$ (Minimax)}}
\psfrag{data5}{\footnotesize{$n = 202$ (c2p)}}
\psfrag{data6}{\footnotesize{$n = 202$ (Minimax)}}
\psfrag{data7}{\footnotesize{$n = 811$ (c2p)}}
\psfrag{data8}{\footnotesize{$n = 811$ (Minimax)}}
\epsfig{width=\figurewidth,file=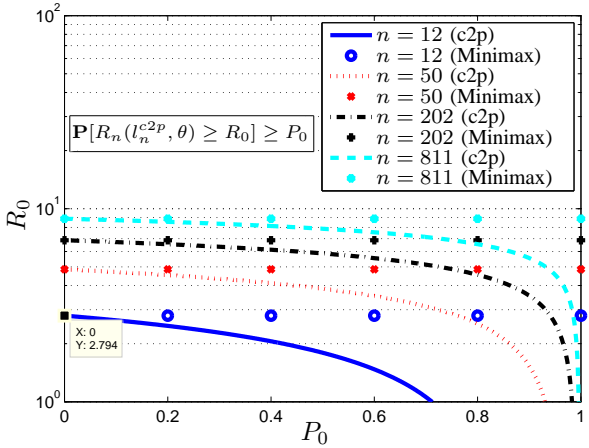}
\vspace{-0.15in}
\vspace{-5pt}
\caption{\small{Average redundancy of the conditional two--stage codes ({\em c2p}) and the average minimax redundancy ({\em Minimax}) as a function of the fraction of sources $P_0$ with $R_n(l_n^{c2p},\theta)>R_0$. First-order Markov source $\mc{M}_1^2$ with $k=2$ and $d=2$.}}
\vspace{-15pt}
\label{fig:M1_2}
\end{figure}

\subsection{Redundancy in Finite-Length Sequences with Small $d$}
\label{subsec:performance_small}

In Figures~\ref{fig:M0_3}~and~\ref{fig:M1_2}, the $x$-axis denotes a fraction $P_0$ and the $y$-axis represents a redundancy level $R_0$. The solid curves demonstrate the derived lower bound on the average redundancy of the conditional two--stage codes $R_0$ as a function of the fraction $P_0$ of the sources with redundancy larger than $R_0$, i.e., $\mb{P}[R_n(l_n^{c2p}, \theta) \geq  R_0] \geq P_0$. In other words, the pair $(R_0, P_0)$ on the redundancy curve means that at least a fraction $P_0$ of the sources that are chosen from Jeffreys' prior have an expected redundancy that is greater than $R_0$.
Note that the unknown parameter vector is chosen using Jeffreys' prior.

\begin{figure}[tb]
\centering
\vspace{-0.05in}
\psfrag{ylabel}{$R_0$}
\psfrag{xlabel}{$P_0$}
\psfrag{dataaaaaaaaaaaaaaaaaaaaaaaa1}{\footnotesize{$n = 4096$ (Two--stage)}}
\psfrag{data2}{\footnotesize{$n = 4096$ (Cond. two--stage)}}
\psfrag{data3}{\footnotesize{$n = 512$ (Two--stage)}}
\psfrag{data4}{\footnotesize{$n = 512$ (Cond. two--stage)}}
\psfrag{data5}{\footnotesize{$n = 64$ (Two--stage)}}
\psfrag{data6}{\footnotesize{$n = 64$ (Cond. two--stage)}}
\psfrag{data7}{\footnotesize{$n = 8$ (Two--stage)}}
\psfrag{data8}{\footnotesize{$n = 8$ (Cond. two--stage)}}
\epsfig{width=\figurewidth,file=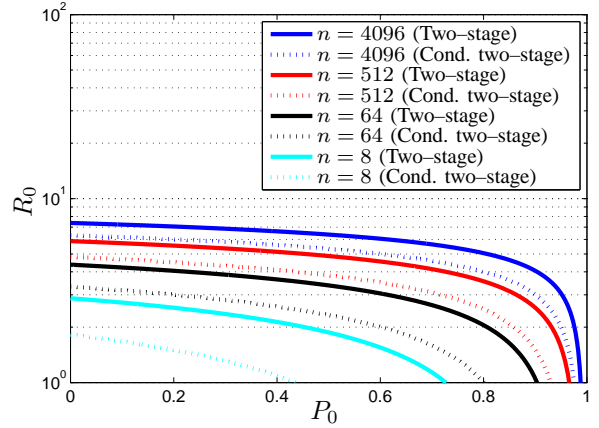}
\vspace{-0.15in}
\vspace{-5pt}
\caption{\small{Average redundancy of the two--stage codes (solid) vs average redundancy of the conditional two---stage codes (dotted) as a function of the fraction of sources $P_0$. Memoryless source $\mc{M}_0^2$ with $k=2$ and $d=1$.}}
\vspace{-15pt}
\label{fig:comparison}
\end{figure}

First, we consider a ternary memoryless information source denoted by $\mc{M}_0^3$. Let $k$ be the alphabet size, where $k = 3$. This source may be parameterized using two parameters, i.e., $d=2$.
In Fig.~\ref{fig:M0_3}, our results are compared to the average minimax redundancy, i.e., $\bar{R}_n$ from~(\ref{eq:minimax}). Since the conditional two--stage codes achieve the minimax redundancy, $\bar{R}_n$ is in fact the average minimax redundancy for the conditional two--stage codes ($\bar{R}_n^{c2p}$) as well. The results are presented in bits.
As shown in Fig.~\ref{fig:M0_3}, at least $40\%$ of ternary memoryless sequences of length $n=32$ ($n=128$) may not be compressed beyond a redundancy of $4.26$ ($6.26$) bits. Also, at least $60\%$ of ternary memoryless sequences of length $n=32$ ($n=128$) may not be compressed beyond a redundancy of $3.67$ ($5.68$) bits. %
Note that as $n\to\infty$, the average redundancy approaches the average minimax redundancy for most sources. 

Next, let $\mc{M}_1^2$ denote a binary first-order Markov source ($d=2$). We present the finite-length compression results in Fig.~\ref{fig:M1_2} for different values of sequence length $n$. The values of $n$ are chosen such that they are almost $\log(3)$ times the values of $n$ for the ternary memoryless source in the first example. This choice has been made to equate the amount of information in the two sequences from $\mc{M}_0^3$ and $\mc{M}_1^2$ allowing a fair comparison. 

Figure~\ref{fig:M1_2} shows that the average minimax redundancy of the conditional two--stage codes for the case of $n = 12$ is given as $\bar{R}_{12} \approx 2.794$ bits. Comparing Fig.~\ref{fig:M0_3} with Fig.~\ref{fig:M1_2}, we conclude that the average redundancy of universal compression for a binary first-order Markov source is very similar to that of the ternary memoryless source, suggesting that $d$ is the most important parameter in determining the redundancy of finite-length sources. This subtle difference becomes even more negligible as $n \to \infty$ since the dominating factor of redundancy for both cases approaches to $\frac{d}{2}\log n$.

As demonstrated in Figs.~\ref{fig:M0_3} and \ref{fig:M1_2}, there is a significant gap between the known result by the average minimax redundancy and the finite-length results obtained in this paper when a high fraction $P_0$ of the sources is concerned. The bounds derived in this paper are tight, and hence, for many sources the average minimax redundancy overestimates the average redundancy in universal source coding of finite-length sequences where the number of the parameters is small. In other words, the compression performance of a high fraction of finite-length sources would be better than the estimate given by the average minimax redundancy.

\begin{figure}[tb]
\centering
\vspace{-0.05in}
\psfrag{ylabel}{$R_0$}
\psfrag{xlabel}{$P_0$}
\psfrag{8E4        }{\footnotesize{$8\times10^4$}}
\psfrag{8E5        }{\footnotesize{$8\times10^5$}}
\psfrag{8E6        }{\footnotesize{$8\times10^6$}}
\psfrag{dataaaaaaaaaaaaaaaaaaaa1}{\footnotesize{$n = 256$kB (c2p)}}
\psfrag{data2}{\footnotesize{$n = 256$kB (Minimax)}}
\psfrag{data3}{\footnotesize{$n = 2$MB (c2p)}}
\psfrag{data4}{\footnotesize{$n = 2$MB (Minimax)}}
\psfrag{data5}{\footnotesize{$n = 16$MB (c2p)}}
\psfrag{data6}{\footnotesize{$n = 16$MB (Minimax)}}
\psfrag{data7}{\footnotesize{$n = 128$MB (c2p)}}
\psfrag{data8}{\footnotesize{$n = 128$MB (Minimax)}}
\epsfig{width= \figurewidth,file=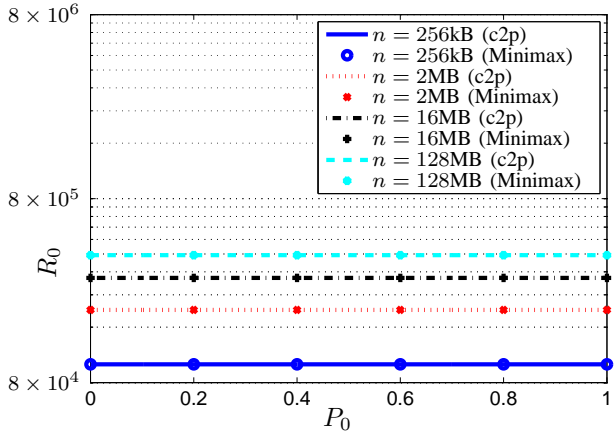}
\vspace{-0.12in}
\vspace{-5pt}
\caption{\small{Average redundancy of the conditional two--stage codes ({\em c2p}) and the average minimax redundancy ({\em Minimax}) as a function of the fraction of sources $P_0$ with $R_n(l_n^{c2p},\theta)>R_0$. First-order Markov source with $k=256$ and $d=65280$. The sequence length $n$ is measured in bytes ($B$).}}
\vspace{-15pt}
\label{fig:M1_256}
\end{figure}

\subsection{Two--Stage Codes Vs Conditional Two--Stage Codes}
\label{subsec:comparison}
We now compare the finite-length performance of the two--stage codes with the conditional two--stage codes on the class of binary memoryless source $\mc{M}_0^2$ with $k=2$ ($d=1$). The results are presented in Figure~\ref{fig:comparison}.
The solid line and the dotted line demonstrate the lower bound for the two--stage codes and the conditional two--stage codes, respectively.
As can be seen, the gap between the achievable compression using two--stage codes and that of the conditional two--stage codes constitutes a significant fraction of the average redundancy for small $n$.
For a Bernoulli source, the average minimax redundancy of the two--stage code is given in~(\ref{eq:minimax_two_part}) as
\begin{equation}
\bar{R}^{2p}_{n} = \bar{R}_n + \frac{1}{2} \log\left(\frac{\pi e}{2} \right)\approx \bar{R}_n + 1.048.
\end{equation}
The average minimax redundancy of two--stage codes for the case of $n = 8$ is given as $\bar{R}^{2p}_8 \approx 2.86$ bits while that of the conditional two--stage codes is $\bar{R}_8 \approx 1.82$. Thus, the two--stage codes incur an extra compression overhead of more than $50\%$ for $n=8$.

In Theorem~\ref{thm:minimax}, we derived that the extra redundancy $g(d)$ incurred by the two--stage assumption.
We further use Stirling's approximation for sources with large number of parameters in order to show the asymptotic behavior of $g(d)$ as $d\to\infty$. That is, asymptotically, we have
\begin{equation}
g(d) = \frac{1}{2}\log\left(\pi d\right) + o(1).
\end{equation}
Note that $o(1)$ denotes a function of $d$ and not $n$ here.
Finally, we must note that the main term of redundancy in  $\bar{R}_n$ is $\frac{d}{2} \log n$, which is linear in $d$, but the penalty term $g(d)$ is logarithmic in $d$. Hence, the effect of the two--stage assumption becomes negligible for the families of sources with larger $d$.

\subsection{Redundancy in Finite-Length Sequences with Large $d$}
\label{subsec:performance_large}
The results of this paper can be used to quantify the significance of redundancy in finite-length compression. We consider a first-order Markov source with alphabet size $k=256$. We intentionally picked this alphabet size as it is a common practice to use the byte as a source symbol. This source may be represented using $d = 256 \times 255 = 62580$ parameters.
 In Figure~\ref{fig:M1_256}, the achievable redundancy is demonstrated for four different values of $n$. Here, again the redundancy is measured in bits. The curves are almost flat when $d$ and $n$ are very large validating our results that the average minimax redundancy provides a good estimate on the achievable compression for most sources. The sequence length in this example is presented in bytes ($B$).
We observe that for $n=256kB$, we have
$R_n(l_n,\theta) \geq 100,000$ bits for most sources.
Further, the extra redundancy due to the two--stage coding $g(d) \approx 8.8$~bits, which is a negligible fraction of the redundancy of $100,000$~bits.
If the source has an entropy rate of $1$ bit per source symbol (byte),  the compression overhead is $38\%$ and $1.7\%$ for sequences of lengths $256$kB and $16$MB, respectively.
Hence, we conclude that redundancy may be significant for the compression of small low entropy sequences. On the other hand, redundancy is negligible for sequences of higher lengths.

\bibliographystyle{IEEEtran}
\bibliography{compress_LB_ISIT11}

\end{document}